\documentclass[prl,twocolumn,showpacs]{revtex4}

\usepackage{graphicx}
\usepackage{soul}
\usepackage{color}

\begin{document}

\title{Gain-Driven Discrete Breathers in ${\cal PT}-$Symmetric Nonlinear Metamaterials}
\author{N. Lazarides$^{1,2}$, G. P. Tsironis$^{1,2}$}
\affiliation{
$^{1}$Department of Physics, University of Crete, P. O. Box 2208, 71003 Heraklion, 
Greece; \\
$^{2}$Institute of Electronic Structure and Laser,
Foundation for Research and Technology-Hellas, P.O. Box 1527, 71110 Heraklion, Greece
}
\date{\today}

\begin{abstract}
We introduce a one dimensional parity-time (${\cal PT}$)-symmetric nonlinear magnetic 
metamaterial consisting of split-ring dimers having both gain and loss. When 
nonlinearity is absent we find a transition between an exact to a broken 
${\cal PT}$-phase; 
in the former the system features a two band gapped spectrum with shape determined by 
the gain and loss coefficients as well as the inter-unit coupling.  
In the  presence of nonlinearity we show numerically that as a result of the 
gain/dissipation matching a novel type of long-lived stable discrete breathers can 
form below the 
lower branch of the band with no attenuation. In these localized modes the energy 
is almost equally partitioned between two adjacent split rings on the one with gain
and the other one with loss.
\end{abstract}


\maketitle

Considerable research effort has recently focused in the investigation and 
developement of artificial materials that exhibit properties
not found in nature. In the electromagnetic domain, these advances resulted in the 
construction of metamaterials, novel artificial structures that provide
full access to all four quadrants of the real permittivity - permeability plane 
\cite{Zheludev2010}.
Recently, there has been increasing interest in synthetic materials with 
a combined parity - time (${\cal PT}$) symmetry. Although quantum systems described 
by ${\cal PT}-$symmetric Hamiltonians have been studied for many years 
(\cite{Hook2012} and refs. therein),
it was only recently realized that many classical systems are 
${\cal PT}-$symmetric \cite{Bender2007}. Subsequently, the notion of ${\cal PT}$ 
symmetry has been extended to dynamical lattices, particularly in optics
\cite{ElGanainy2007,Makris2008}. Soon after that, ${\cal PT}-$symmetry breaking
was experimentally observed \cite{Guo2009,Ruter2010,Szameit2011}.
Such considerations have been also extended in nonlinear lattices, where the 
existence of stable discrete solitons \cite{Dmitriev2010}
and Talbot effects \cite{Ramezani2012} was theoretically demonstrated.

Among recent developements in ${\cal PT}-$symmetric materials, the application
of these ideas in electronic circuits \cite{Schindler2011} not only provides a 
platform for testing 
the new ideas within the framework of easily accessible experimental configurations,
but also provides a link to metamaterials.
Conventional, metallic metamaterials suffer from high losses that hamper their
use in practical applications. 
However, building metamaterials with ${\cal PT}$ symmetry, relying on gain and loss,
may provide a way out and moreover lead to new extraordinary properties.
It is shown that ${\cal PT}-$symmetric metamaterials undergo spontaneous symmetry 
breaking from the exact ${\cal PT}$ phase (real eigenfrequencies) to the broken 
${\cal PT}$ phase (at least a pair of complex eigenfrequencies), with variation of the 
gain/loss coefficient. In the presence of nonlinearity, the generation of 
long-lived excitations in the form of discrete breathers (DBs) \cite{Flach2008} 
is demonstrated numerically in metamaterial models.
These novel gain-driven DBs, generated either by proper intialization of the 
${\cal PT}$ metamaterial or purely dynamically through external driving,
result from power matching of the input power through the gain mechanism and
internal loss.
\begin{figure}[!h]
\includegraphics[angle=0, width=0.8 \linewidth]{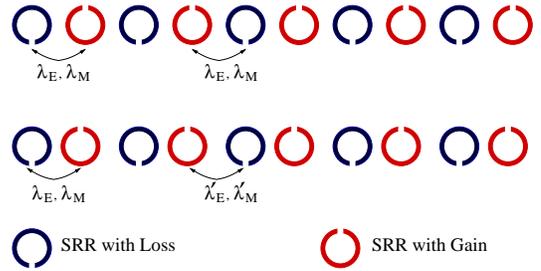}
\caption{(Color online)
Schematic of a ${\cal PT}$ metamaterial.
Upper panel: all the SRRs are equidistant.
Lower panel: the separation between SRRs is modulated according to a binary pattern
(${\cal PT}$ dimer chain).
}
\end{figure}

Consider a one-dimensional array of dimers, each comprising two nonlinear
split-ring resonators (SRRs); one with loss and the other with equal amount 
of gain (Fig. 1). The SRRs are coupled magnetically and/or electrically 
through
dipole-dipole forces \cite{Sydoruk2006,Hesmer2007,Sersic2009,Rosanov2011}
and are regarded as RLC circuits, featuring a resistance $R$,
an inductance $L$, and a capacitance $C$. Early realizations of nonlinear
metamaterial elements employed a varicap diode \cite{Reynet2004};
subsequently several types of diodes have been employed to demonstrate
single tunable elements \cite{Powell2007,Wang2008} and metamaterials 
\cite{Shadrivov2008}.
For constructing  active metamaterials, the incorporation of
constituents that provide gain through external energy sources is a 
promising technique \cite{Boardman2010}. In particular 
realizations, a tunnel (Esaki) diode \cite{Esaki1958} featuring negative
resistance has been employed \cite{Jiang2011}.  
Low-loss, active metamaterials have been demonstrated
as left-handed transmission lines \cite{Jiang2011} and optical fishnet 
structures \cite{Xiao2010}. However, there are still open issues concerning
gain operation in metamaterials, related to noise and other effects that 
come into play \cite{Syms2011}.

In the equivalent circuit model picture 
\cite{Lazarides2006,Molina2009,Lazarides2011,Rosanov2011},
extended for the ${\cal PT}$ dimer chain, the dynamics of the charge $q_n$
in the capacitor of the $n$th SRR is governed by 
\begin{eqnarray}
\label{1}
   \lambda_M' \ddot{q}_{2n} &+&\ddot{q}_{2n+1} +\lambda_M \ddot{q}_{2n+2}
  +\lambda_E' q_{2n} +q_{2n+1} +\lambda_E q_{2n+2}
 \nonumber \\
   &=&\varepsilon_0 \sin(\Omega \tau) 
   -\alpha q_{2n+1}^2 -\beta q_{2n+1}^3 -\gamma \dot{q}_{2n+1} \\ 
\label{2}
   \lambda_M \ddot{q}_{2n-1} &+&\ddot{q}_{2n} +\lambda_M' \ddot{q}_{2n+1}
   +\lambda_E q_{2n-1} +q_{2n} +\lambda_E' q_{2n+1} 
 \nonumber \\
    &=&\varepsilon_0 \sin(\Omega \tau)
    -\alpha {q}_{2n}^2 -\beta {q}_{2n}^3 +\gamma \dot{q}_{2n}  
\end{eqnarray}
where 
$\lambda_M, \lambda_M'$ and $\lambda_E, \lambda_E'$ are the magnetic and electric 
interaction coefficients, respectively, between nearest neighbors, 
$\alpha$ and $\beta$ are nonlinear coefficients,
$\gamma$ is the gain/loss coefficient ($\gamma >0$), 
$\varepsilon_0$ is the amplitude of the external driving voltage,
while $\Omega$ and $\tau$ are the driving frequency and temporal variable,
respectively, normalized to $\omega_0 =1/\sqrt{L C_0}$ and $\omega_0^{-1}$, 
respectively, with $C_0$ being the linear capacitance.
We have also considered additional next-nearest neighbor coupling between SRRs,
with coefficients that fall off as the inverse-cube of the distance between
them. Although slight quantitative changes are observed, the  
results presented below are not qualitatively affected.   
The chosen values of the nonlinearity coefficients are typical for a diode 
\cite{Lazarides2011}, while $\gamma$ chosen to provide stable operation.
The coupling coefficients are chosen relatively large for clarity. 
However, breathers appear generically even for weakly coupled SRRs.

By substituting $q_{2n} = A \exp[i( 2 n \kappa -\Omega \tau)]$ and 
$q_{2n+1}=B \exp[i( (2 n+1) \kappa -\Omega \tau)]$, 
where $\kappa$ is the normalized wavevector, into Eqs. (\ref{1}) and (\ref{2}),
and requesting nontrivial solutions for the resulting stationary problem,
we obtain
\begin{eqnarray}
 \label{5}
  \Omega_\kappa^2 =\left(  -b \pm \sqrt{\Delta} \right)/(2 a) ,
\end{eqnarray}
where $a= 1 -(\lambda_M -\lambda_M')^2 -\mu_\kappa \mu_\kappa'$,
$b= \gamma^2 -2 \left[ 1 -(\lambda_E -\lambda_E') (\lambda_M -\lambda_M') \right]
                 +\varepsilon_\kappa \mu_\kappa' +\varepsilon_\kappa' \mu_\kappa$,
$c= 1 -(\lambda_E -\lambda_E')^2 -\varepsilon_\kappa \varepsilon_\kappa'$, 
$\Delta =b^2 -4 a c$, and $\varepsilon_\kappa = 2 \lambda_E \cos(\kappa)$,
$\varepsilon_\kappa' = 2 \lambda_E' \cos(\kappa)$,
$\mu_\kappa = 2 \lambda_M \cos(\kappa)$, $\mu_\kappa' = 2 \lambda_M' \cos(\kappa)$. 
In the following, we consider that the relative orientation of the SRRs in the chain
is such that the magnetic coupling dominates, while the electric coupling can be 
neglected (Fig. 1) \cite{Hesmer2007}. Then, Eq. (\ref{5}) reduces to 
\begin{eqnarray}
 \label{9}
  \Omega_\kappa^2 = \frac{2-\gamma^2 \pm \sqrt{\gamma^4 -2 \gamma^2 
               +(\lambda_M -\lambda_M')^2 +\mu_\kappa \mu_\kappa'} }
             {2 (1 -(\lambda_M -\lambda_M')^2 -\mu_\kappa \mu_\kappa')} .
\end{eqnarray}
\begin{figure}[!t]
\includegraphics[angle=0, width=0.75 \linewidth]{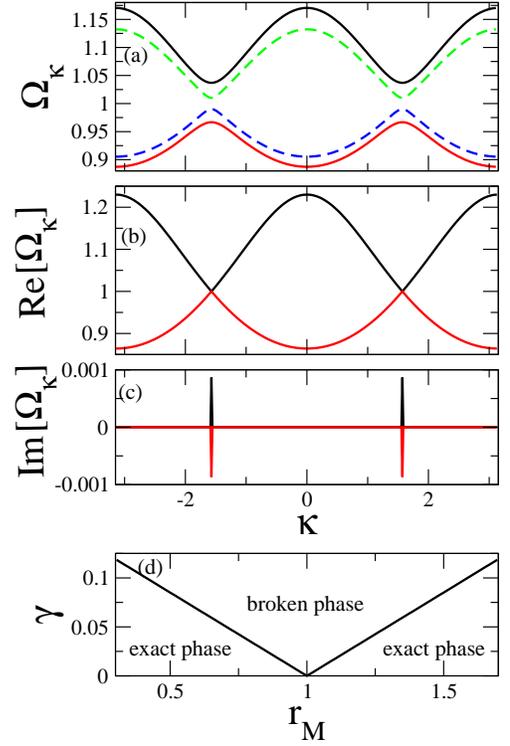}
\caption{(Color online)
(a) Frequency bands for a ${\cal PT}$ metamaterial for $\gamma=0.002$,
and
$\lambda_M=-0.17$, $\lambda_M' =-0.10$
(black and red solid lines);  
$\lambda_M =-0.12$, $\lambda_M' =-0.10$ (green and blue dashed lines). 
The imaginary parts are zero.
(b) $\&$ (c) Real and imaginary parts, respectively, of the frequency bands 
for a ${\cal PT}$  metamaterial with $\gamma=0.002$, $\lambda_M=-0.17$,
$\lambda_M' =-0.1699$.
(d) ${\cal PT}$ phase diagram on the 
$\gamma -r_M$ plane ($r_M =\lambda_M' /\lambda_M$), for $\lambda_M=-0.17$. 
}
\end{figure}
The condition for $\Omega_\kappa$ being real for any $\kappa$ then reads
\begin{eqnarray}
\label{10}
  \cos^2 (\kappa) \geq \frac{\gamma^2 (2 -\gamma^2) -(\lambda_M -\lambda_M')^2}
                            {4 \lambda_M \lambda_M'} .
\end{eqnarray}
For $\lambda_M =\lambda_M'$ Eq. (\ref{10}) cannot be satisfied for all 
$\kappa$'s for any $\gamma >0$, implying that a large ${\cal PT}-$symmetric
SRR array (Fig. 1, upper) is in the broken phase. 
However, for a dimer chain with $\lambda_M \neq \lambda_M'$
(Fig. 1, lower), the above condition is satisfied for all $\kappa$'s for 
$\gamma \leq \gamma_c \simeq |\lambda_M -\lambda_M'|$,  ($\gamma^4 \simeq 0$). 
In the exact phase ($\gamma < \gamma_c$), the ${\cal PT}$ dimer chain 
has a gapped spectrum with two frequency bands [Fig 2(a)]. 
The width of the gap separating the bands decreases with decreasing
$|\lambda_M -\lambda_M'|$ for constant $\gamma$. For $\gamma \simeq \gamma_c$
the gap closes [Fig. 2(b)], some frequencies in the spectrum acquire 
imaginary parts [Fig. 2(c)] and ${\cal PT}-$symmetry breaking occurs. 
A typical phase diagram on the $\gamma - r_M$ plane,
with $r_M = \lambda_M' /\lambda_M$, is shown in Fig. 2(d). The solid lines indicate
the variation of the exceptional point $\gamma_c$ that separates the exact
from the broken phase. 
For fixed $\lambda_M, \lambda_M'$, the bandwidths as a function of $\gamma$ are
shown in Fig. 3.
\begin{figure}[!t]
\includegraphics[angle=0, width=0.8 \linewidth]{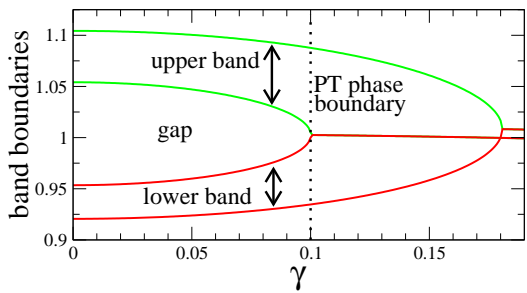}
\caption{(Color online)
Frequency band boundaries as a function of $\gamma$
for a ${\cal PT}$ dimer chain with $\lambda_M=-0.14$, $\lambda_M' =-0.04$.
}
\end{figure}

Eqs. (\ref{1}) and (\ref{2}), implemented with $q_0 (\tau) = q_{N+1} (\tau) =0$,
are integrated numerically with
$q_m (0) =(-1)^{m-1} {\rm sech}(m/2)$, $\dot{q}_m (0) =0$,  
and $\varepsilon_0 =0$, $\alpha=-0.4$, $\beta=0.08$.
Using a gain/loss function of the form
\begin{equation}
 \label{12}
  \gamma (m) = \left \{ \begin{array}{lll}
      \gamma, & m=1,...,N_\ell \\
      (-1)^{m-1} \gamma, & m=N_\ell+1,...,N-N_\ell \\
      \gamma, & m=N-N_\ell+1,...,N ,
\end{array}
\right.
\end{equation}
we obtain gain-driven, long-lived ($>10^8$ time units) DBs for wide parameter
intervals.
Note that the actual ${\cal PT}$ dimer chain is of length $N-2N_\ell$,
while both its ends are joined to lossy dimer chains of length $N_\ell$.
We found empirically that this is the most effective way to stabilize DBs 
in the ${\cal PT}$ metamaterial, i.e., by embedding it into a lossy metamaterial.
The lossy parts help the excess energy to go smothly away during the long transient 
phase of integration, and thus prevents the blowing up of the solution that 
otherwise may have occured.
The energy density $E_n$ evolution of a typical DB in the $n-\tau$ plane
is shown in Fig. 4. The largest part of the total energy 
is concentrated into two neighboring sites belonging to the same dimer.
The corresponding instantaneous current profile $i_n$ at maximum current
(Fig. 5) is neither symmetric or antisymmetric at the single SRR level.
\begin{figure}[!t]
\includegraphics[angle=0, width=0.8 \linewidth]{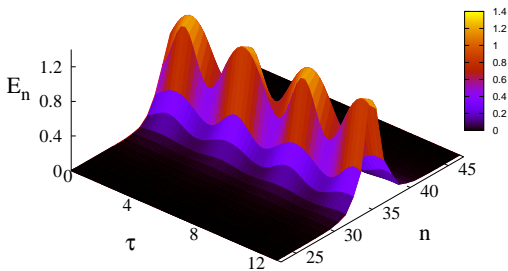} 
\caption{(Color online)
Spatiotemporal evolution of the energy density $E_n$ during two periods
for $N=70$, $N_\ell =10$, $\lambda_M' =-0.10$, $\lambda_E =\lambda_E' =0$, 
$\gamma=0.002$, and $\lambda_M=-0.17$. 
}
\end{figure}
\begin{figure}[!t]
\includegraphics[angle=0, width=0.8 \linewidth]{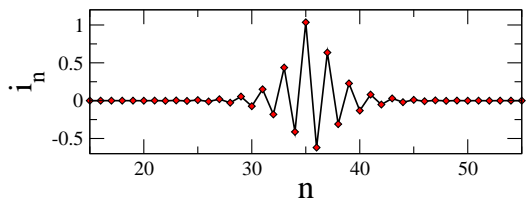}
\caption{(Color online)
Gain-driven, current breather profile $i_n$ as a function of $n$, for 
$N=70$, $N_\ell =10$, $\lambda_M' =-0.10$, $\lambda_E =\lambda_E' =0$, 
$\gamma=0.002$, and $\lambda_M=-0.17$.
}
\end{figure}

The breather frequency $\Omega_B$ can be obtained from the power 
spectrum of a time series of the energy in one of the DB sites.
The logarithm of a typical power spectrum is plotted as a function of frequency
in Fig. 6(a); strong harmonics to the fundamental frequency $\Omega_B$ are observed.
The frequency $\Omega_B$ increases with decreasing $|\lambda_M -\lambda_M'|$
[Fig. 6(b)], while the variation of  
$\gamma$ within $10^{-2} - 10^{-3}$ has apparently no effect on $\Omega_B$
at least in the particular case shown in Fig. 6(c).
The dependence of $\Omega_B$ on $|\lambda_M -\lambda_M'|$
is illustrated in Fig. 6(d), for those values of $|\lambda_M -\lambda_M'|$ 
for which DBs are stable (or at least long-lived). 
\begin{figure}[!t]
\includegraphics[angle=0, width=0.8 \linewidth]{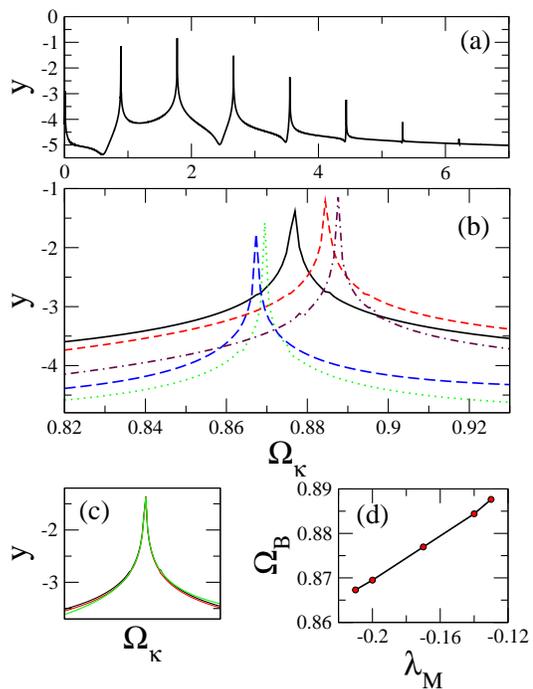}
\caption{(Color online)
(a) The logarithm of the power spectrum of energy at one of the central 
breather sites $y=log_{10}[PS(E_{n=N/2})]$ as a function of $\Omega_\kappa$, 
for  
$N=70$, $\lambda_M' =-0.10$, $\lambda_M=-0.13$, $\lambda_E =\lambda_E' =0$, 
$\gamma=0.002$.
(b) The same as in (a) around the fundamental breather frequency $\Omega_B$,
for $\lambda_M=-0.21$ (blue - long dashed); 
$\lambda_M=-0.20$ (green - dotted); $\lambda_M=-0.17$ (black - solid);
$\lambda_M=-0.14$ (red - dashed); $\lambda_M=-0.13$ (maroon - dotted dashed). 
The other parameters as in (a).
(c) The same as in (a) and (b) around $\Omega_B$,
for $\gamma=0.002$ (black); $0.005$ (green); $0.01$ (red). 
(d) Dependence of $\Omega_B$ on $\lambda_M$ for the spectra in (b).
}
\end{figure}

For a gapped linear spectrum, large amplitude linear modes become unstable
in the presence of driving and nonlinearity. If the curvature of the dispersion
curve in the region of such a mode is positive and the lattice potential is
soft, large amplitude modes become unstable with respect to DB formation
in the gap below the linear spectrum \cite{Sato2003}.
For the parameters of Fig. 2(a), the bottom of the lower band is located 
at $\Omega_0 \simeq 0.887$, where the curvature is positive.
Moreover, the SRRs are subjected to soft on-site potentials for the selected 
values of $\alpha$ and $\beta$.
Then, DBs can be generated spontaneously by a frequency chirped driver as it is
illustrated in Fig. 7, where the energy density $E_n$ 
for a ${\cal PT}$ dimer chain is plotted on the $n-\tau$ plane.
We use the following procedure:
At time $\tau=0$, we start integrating Eqs. (\ref{1}) and (\ref{2}) with zero
initial state and external driving for $500~T_0 \simeq 3500$ time units (t.u.), 
where $T_0 =2\pi/\Omega_0$,
to allow for significant developement of large amplitude modes. 
At time $\tau \simeq 3500$ t.u. (point A on Fig. 7), the driver is switched-on
with low-amplitude and frequency slightly above $\Omega_0$
($1.01~\Omega_0 \simeq 0.894$). The frequency is then chirped downwards with 
time to induce instability for the next $10,600$ t.u. ($\sim 1500~T_0$), 
until it is well below $\Omega_0$ ($0.997~\Omega_0 \simeq 0.882$).
During that phase, a large number of excitations are generated that move and 
strongly interact to each other, eventually merging into a small number of high 
amplitude (multi-)breathers.
At time $\tau \simeq 14,100$ t.u. (point B on Fig. 7), the driver is switched off
and the DBs that have formed are solely driven by the gain. They continue to
interact until they reach a stationary state and get trapped at particular sites.   
The high density segments between points B and C in Fig. 7 present precisely
those stationary gain-driven  (multi-)breathers generated through 
chirping and subsequent dynamics.
At time $\tau \sim 85150$ t.u. (point C on Fig. 7), the gain is replaced by equal
amount of loss, and the breathers die out rapidly.  
\begin{figure}[!t]
\vspace{-3cm}
\includegraphics[angle=0, width=1.4 \linewidth]{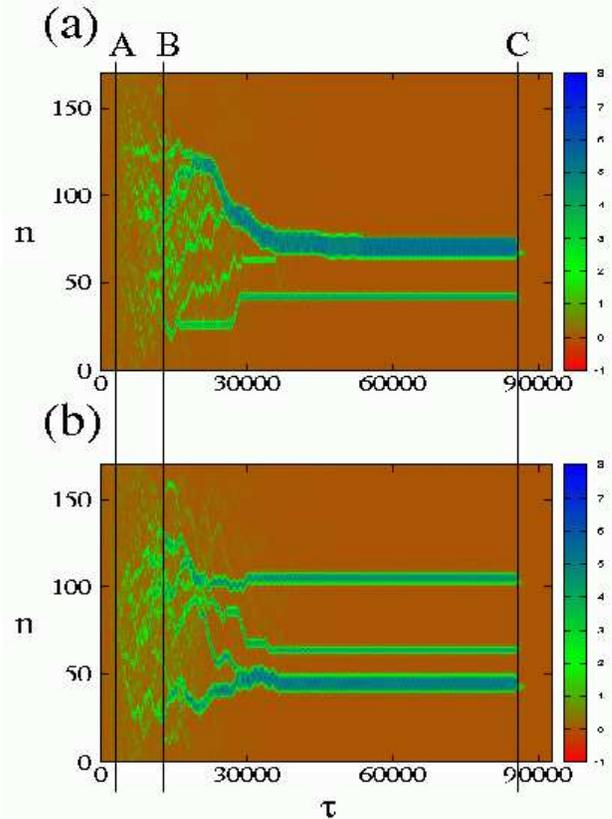}
\caption{(Color online)
The energy density $E_n$ on the $n-\tau$ plane for a ${\cal PT}$
dimer chain with $N=170$, $N_\ell =10$, $\Omega_0 =0.887$, $\gamma=0.002$,
$\lambda_M =-0.17$, $\lambda_M' =-0.10$ ($\lambda_E =\lambda_E' =0$), and 
(a) $\varepsilon_0 =0.085$; (b) $\varepsilon_0 =0.095$.
The vertical lines separate different stages in the chirping procedure.
}
\end{figure}

The construction of a ${\cal PT}$ metamaterial is feasible with the present 
technology in the microwaves, where negative resistance devices
\cite{Boardman2007,Jiang2011} and transistors \cite{Xu2012} are convenient 
for providing gain. Thus, ${\cal PT}$ dimers
can be constructed and balanced in a way similar to that in electrical 
circuits \cite{Schindler2011}. For SRR dimers, highly conducting rings are 
required for reduced Ohmic losses. The bias of the negative resistance device 
in the SRR with gain should be adjusted to provide a gain coefficient with
magnitude equal to that of the loss coefficient of the SRR without gain.
A chain of such dimers makes a ${\cal PT}$ metamaterial whose dynamics
is approximatelly described by model Eqs. (\ref{1}) and (\ref{2}).

For a balanced configuration the metamaterial exhibits two bands separated
by a gap. While the 
band structure appears to be stable to small structural perturbations, it may 
be more fragile in the case of relatively strong disorder 
\cite{Bendix2009,Lazarides2012}.
However, any divergence related to a disorder-induced broken phase will manifest
itself in time-scales larger than the characteristic interdimer dynamics
time-scale, making possible the experimental observation in real ${\cal PT}$ 
systems (e.g. in Ref. \cite{Schindler2011}).

It is demonstrated numerically that stable or at least
long-lived, gain-driven DBs may be excited generically.
DBs result either by proper initialization of the system, or purely dynamically
by frequency chirping of a weak alternating driver. 
Fundamental DBs occupy two neighboring sites that belong to the same
dimer. Stability and long life-times of DBs is achieved by embedding the 
${\cal PT}-$symmetric dimer chain into a lossy dimer chain.
When the balance between gain and loss is not exact, i.e., when the magnitudes of
the gain and loss coefficients differ by a small amount, DBs can still be 
generated through the chirping procedure. In this case, however, the DBs 
lose their long-term stability, viz. for loss exceeding gain they decay
slowly until they vanish while in the opposite case the DBs gain energy and
diverge. 
Both effects occur on a time-scale related to the gain-loss imbalance. 


This work was supported by the "THALES" Project ANEMOS,
funded by the ESPA Program of the Ministry of Education of Greece.


\end{document}